# A Note on Relative Consciousness

Kynan Eng

Institute of Neuroinformatics, University of Zurich and ETH Zurich

## Abstract

This paper describes a mathematical formulation for measuring how one system can estimate the consciousness of another. This consciousness estimate is always relative to the observer. The paper shows how this formulation leads to simple resolutions of some key problems of consciousness.

## Introduction

Many discussions of the concept of consciousness implicitly or explicitly assume the existence of consciousness as a real "quantity" or "thing". One form that this belief takes is in the idea of panpsychism (Chalmers 2015), in which all fundamental particles have a mental state. A somewhat weaker or agnostic version of this belief postulates that while consciousness is possibly a "real" thing, it is hard (or impossible?) to be measured directly. However, supporters of this idea suggest that it is possible to gain an approximate measure of consciousness by correlative means, i.e. neural correlates of consciousness (NCC) (Chalmers 2000).

The idea of consciousness as a real thing, whether pervasive or not directly measurable, may have some echoes in the long-abandoned search for the luminiferous æther for electromagnetic phenomena (Dirac 1951). Current attempts to quantify consciousness also suffer from problems in achieving universality, in particular in encompassing many levels of system complexity (for example, humans versus insects).

I propose that the attempt to find a universal NCC or panpsychic definition of consciousness is either ill-placed or doomed to fail. Instead, we can consider an alternative way of thinking about consciousness with the following base assumptions:

- Any measurement of consciousness must occur with a nominated observer and subject. There is no such thing as absolute consciousness without either of these two elements.
- Any measure of consciousness is always in the frame of reference of the observer. There is no absolute observer.
- Any measurement of consciousness occurs at a certain point in time. It is valid only for that measurement, and may change over time as the observer evolves.

The rest of this note concerns a formulation of these principles, by synthesizing some concepts of information transmission and complexity. This is followed by some commentary on what these principles mean in the light of some classical conundrums in consciousness research.

# Formulation

Consider an agent $B$, observing another agent $A$. $B$ wishes to ascertain the consciousness of $A$. $B$ observes $A$, and thus $B$ receives a certain amount of information $I_{BA}$ from $A$. We denote the result of this process as $B$'s consciousness of $A$, i.e. $C_{BA}$. Note that this consciousness $C_{BA}$. is local to $B$.

To obtain and internally represent $C_{BA}$, $B$ must have a certain set of attributes:

- an ability to observe $A$, i.e. to receive information transmitted from $A$;
- a level of ability to create and process a model of $A$ within itself, which we shall call a schema $b$;
- (optionally, an ability to perturb $A$, i.e. send information to $A$)

In information theory terms, this is a problem in which $B$ attempts to infer the internal structure of $A$, based on observed information about $A$. $B$ has a certain internal schema b, which is smaller than or equal to $B$, to create and sustain a given level of consciousness of $A$.

We describe the overall process as follows (Figure 1):

$$C_{BA}^b = b(I_{BA})$$

Without loss of generality, $B$ may have more than one schema $b1$, $b2$, ... for creating different estimates of the consciousness of $A$, i.e. $C_{BA}^{b1}, C_{BA}^{b2}, ...$

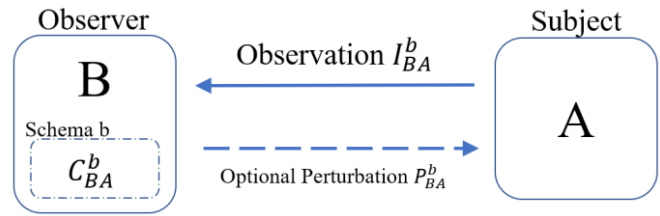

*Figure 1: Relative consciousness observation process.*

It is important to note that the schema $b$ is not just a process or algorithm, but also a set of priors encoded either explicitly or in the algorithm of $b$ itself. This implies that it is entirely possible that if $B$ is using the wrong $b$ and not "looking" at $A$ in the "right" way, it could receive a result of zero.

Note that the observations may be discrete, batched, or time-continuous, i.e. $I_{BA}^b(t)$. Similarly, the schema may also vary with time $b(t)$, leading to time-varying estimates of consciousness $C_{BA}^b(t)$.

Now, given $I_{BA}$, there may be ways to estimate certain bounds on $B$, i.e. the requirements for $B$ to have obtained the information $I_{BA}$. One way to do this, for example, could be to use a complexity measure.

# Corollaries and Predictions

The formulation proposed above leads to a number of corollaries and predictions, including the following:

- The maximum complexity of $C_{BA}$, i.e. the internal consciousness model of $A$ constructed by $B$, is affected by the chosen schema $b$.
- Any system $B$ cannot be fully consciousness of itself, i.e. $C_{BB}$ is not observable, because then $b$ would be equal to or larger than $B$. However, $B$ can be conscious of parts of itself, i.e. $C_{Bd}$ may be observable, where $d \in B$, $d \neq B$.

- Any system *B* cannot be fully conscious of another structurally similar system *B'*, where structural similarity is defined as equivalent complexity. We will define *B'* as a *peer system*. In other words, $C_{BB'}$ is not observable in the general case. However, $C_{BB'}^b$ may be observable, because *B* has selected some simplified schema *b* for its observations (*b*∈*B*, *b*≠*B*).
- Any system *B* which is contained inside a larger system *D* (*B*∈*D*) could be conscious of *D*, as long as the selected schema *b* is sufficiently simple.
- Any system *B* can be conscious of what another system *D* is conscious of when observing *A*, i.e. $C_{BD,DA}^{b,d} = b(d(I_{DA}))$ may exist, as long as *b* can encode *d*. Note that *d* may not be directly observable in *D*. In the limiting case, however, *d* = *D* and $b(D(I_{DA}))$ may exist.

# Discussion

Given the preceding formulation, one can make a number of statements about classical problems in consciousness research:

*That was a trick. $C_{BA}$ says nothing about the consciousness of A.*

That is partly correct, but only if one assumes that A has an absolute consciousness. The present formulation says only a limited amount about the consciousness of *A*, in that it must have been complex enough to encode $I_{BA}$. However, it also says something about the consciousness of *A*, as observed by *B*, and the structure of the schema used by *B*. This is a joint relative consciousness. To borrow a term from quantum physics, the consciousness of *A* could be said to be entangled with its observation by *B*. In everyday language, the following statement could be made: just because one did not understand the level of consciousness of another being, it didn't mean that it wasn't always there. Any newly reported level of consciousness says as much (if not more) about the observer as it does about the subject under observation.

*What about the Chinese Room problem?*

Searle's Chinese Room thought experiment (Searle 1982) asserts that a computer cannot have "strong AI", i.e. a true "mind" in the same way as a human, because it is just following instructions without any true "understanding". What is not usually stated in the formulation of this problem, but is a key hidden assumption, is that a human *H* can understand its own consciousness, i.e. $C_{HH'}$ is observable. We have already seen before that this is not possible in the general case, unless one chooses a simplified schema *h* to observe *H'*. Hence the problem is ill-posed due to an incorrect assumption.

*What about qualia?*

The relative consciousness formulation does not exclude the possibility of transferable qualia (Stubenberg 1998, Jackson 1982), i.e. $C_{BA} = C_{DA}$. However, observing such an equivalence requires another observer that is able to encode both *B* and *D*, at the level of complexity required for both *B* and *D* to make their observations of *A*. This third observer cannot be *B* or *D*, because *B* and *D* alone cannot encode sufficiently complex representations of themselves observing A. Thus *B* or *D* cannot conclude that their qualia are in fact the same thing.

*What about artificial general intelligence?*

The definition does not exclude the possibility of a machine matching human intelligence (which, for the purpose of this discussion, we define as being correspondingly equivalent to consciousness). What it does exclude, however, is the capacity of a human to understand whether or not it is observing a human-equivalent intelligence. See the Chinese Room problem.

*What about panpsychism?*

The definition is applicable to any system, and does not exclude the possibility that an observer might describe significant consciousness in everything that it observes, i.e. $C_{BD}^b > 0$ for all values of $D$. This is most likely to occur where $b$ is chosen to be relatively simple, i.e. looking for "microphenomenal consciousness".

*What about the NCC?*

The NCC requires a modification in its formulation to be a valid approach. Instead of the all-encompassing "neural correlate of consciousness", it should be a "neural correlate of consciousness of $X$". For example, the "NCC of self", the "NCC of the external world", the "NCC of a moving stimulus", the "NCC of a painful stimulus", etc. Sometimes, $X$ is some part of the external world, while other times it is some aspect of the system itself. This approach already seems to be followed implicitly by many researchers; the suggestion here is to make the formulation explicit at all times.

*So is consciousness a real thing, or an arbitrary definition?*

It is both, in the same way that information is both real (based on physical processes) and arbitrary (one of many possible definitions depending on the level of description being used).

*Is there a maximum speed of consciousness?*

Yes, because it is anchored in information, which can in turn be anchored in physical properties.

*Can consciousness be transferred?*

The question is not posed precisely enough. The question should be: can my consciousness of $A$ be transferred to my consciousness of $B$, if I evaluate this consciousness in a certain way? The answer is generally yes, for certain values of "me", $A$, $B$, and how I am looking at $A$ and $B$.

*Can we be conscious of the universe?*

Yes, $C_{BU}^b$ can be valid for a universe $U$, $B \in U$, if one chooses a suitably simple schema $b$.

# Further Work

This note aimed to show a formulation of relative consciousness that could be used across a wide variety of systems. It was also an attempt to cast the discussion of consciousness research in a way that may be within reach of information theory. Future work could investigate the following:

- Information-theoretic proofs related to the presented formulation
- Extending the concept to time series and dynamical systems

# Acknowledgments

Thanks to the initiators of the seminar *Consciousness: From Philosophy to Neuroscience* run at the University of Zurich and ETH Zurich during the early 2000s (Daniel Kiper, Heather Berlin, and Christof Koch), for stimulating my interest in the topic. The concepts in this note date back to some of my thoughts from that period.

# References

Chalmers DJ. What is a neural correlate of consciousness. Neural correlates of consciousness: Empirical and conceptual questions. 2000 Sep:17-39.

Chalmers DJ. Panpsychism and panprotopsychism. Consciousness in the physical world: Perspectives on Russellian monism. 2015 Apr 1:246-76.

Dirac PAM. Is there an Æther? Nature. 1951 168:906-907.

Jackson F. Epiphenomenal qualia. The Philosophical Quarterly (1950-). 1982 Apr 1;32(127):127-36.

Searle JR. The Chinese room revisited. Behavioral and brain sciences. 1982 Jun;5(2):345-8.

Stubenberg L. Consciousness and qualia. John Benjamins Publishing; 1998.